\documentclass[english]{IEEEtran}
\usepackage[T1]{fontenc}
\usepackage[latin9]{inputenc}
\usepackage{amsmath}
\usepackage{amsthm}
\usepackage{amssymb}

\makeatletter
\theoremstyle{plain}
\newtheorem{thm}{\protect\theoremname}
\theoremstyle{plain}
\newtheorem{lem}[thm]{\protect\lemmaname}
\theoremstyle{remark}
\newtheorem{rem}[thm]{\protect\remarkname}
\theoremstyle{definition}
\newtheorem{defn}[thm]{\protect\definitionname}
\theoremstyle{plain}
\newtheorem{cor}[thm]{\protect\corollaryname}
\theoremstyle{definition}
\newtheorem{example}[thm]{\protect\examplename}
\theoremstyle{plain}
\newtheorem{conjecture}[thm]{\protect\conjecturename}

\usepackage{amsmath}
\usepackage{tikz,pgfplots,bbm}
\pgfplotsset{compat=1.12}
\usepackage[keeplastbox]{flushend}

\date{}

\newcommand{\arikan}{Ar{\i}kan}

\IEEEoverridecommandlockouts

\renewcommand{\footnotemark}{}

\makeatother

\usepackage{babel}
\providecommand{\conjecturename}{Conjecture}
\providecommand{\corollaryname}{Corollary}
\providecommand{\definitionname}{Definition}
\providecommand{\examplename}{Example}
\providecommand{\lemmaname}{Lemma}
\providecommand{\remarkname}{Remark}
\providecommand{\theoremname}{Theorem}

\begin{document}

\title{Near-Optimal Finite-Length Scaling for\\
Polar Codes over Large Alphabets}

\author{Henry D. Pfister and R\"{u}diger Urbanke \thanks{This work was presented in part at the 2016 International Symposium on Information Theory (ISIT) in Barcelona, Spain. It was initiated while the authors were visiting the Simons Institute at Berkeley for the 2015 program on Information Theory. The work of H.~D.~Pfister was supported in part by the National Science Foundation (NSF) under Grant No.~1545143.     Any opinions, findings, conclusions, and recommendations expressed in this material are those of the authors and do not necessarily reflect the views of these sponsors.} \thanks{H.~D.~Pfister is with the Department of Electrical and Computer Engineering, Duke University (email: henry.pfister@duke.edu).} \thanks{R.~Urbanke is with the School of Computer and Communication Sciences, EPFL, Switzerland (email: ruediger.urbanke@epfl.ch)}  }
\maketitle
\begin{abstract}
For any prime power $q$, Mori and Tanaka introduced a family of $q$-ary polar codes based on $q$~by~$q$ Reed-Solomon polarization kernels. For transmission over a $q$-ary erasure channel, they also derived a closed-form recursion for the erasure probability of each effective channel. In this paper, we use that expression to analyze the finite-length scaling of these codes on the $q$-ary erasure channel with erasure probability $\epsilon\in(0,1)$. Our primary result is that, for any $\gamma>0$ and $\delta>0$, there is a $q_{0}$ such that, for all $q\geq q_{0}$, the fraction of effective channels with erasure rate at most $N^{-\gamma}$ is at least $1-\epsilon-O(N^{-1/2+\delta})$, where $N=q^{n}$ is the blocklength. Since this fraction cannot be larger than $1-\epsilon-O(N^{-1/2})$, this establishes near-optimal finite-length scaling for this family of codes. Our approach can be seen as an extension of a similar analysis for binary polar codes by Hassani, Alishahi, and Urbanke.

A similar analysis is also considered for $q$-ary polar codes with $m$ by $m$ polarizing matrices. This separates the effect of the alphabet size from the effect of the matrix size. If the polarizing matrix at each stage is drawn independently and uniformly from the set of invertible $m$ by $m$ matrices, then the linear operator associated with the Lyapunov function analysis can be written in closed form. To prove near-optimal scaling for polar codes with fixed $q$ as $m$ increases, however, two technical obstacles remain. Thus, we conclude by stating two concrete mathematical conjectures that, if proven, would imply near-optimal scaling for fixed~$q$.
\end{abstract}


\global\long\def\Pr{\mathbb{P}}

\global\long\def\expt{\mathbb{E}}

\global\long\def\vecnot#1{\underline{#1}}

\global\long\def\supp{\mbox{supp}}

\global\long\def\rank{\mathrm{rk}}

\global\long\def\dd{\mathrm{d}}

\global\long\def\midb#1{\middle#1}

\global\long\def\ind{\mathbbm{1}}

\section{Introduction}

To achieve reliable communication at rates close to the channel capacity, it is well-known that the blocklength must tend to infinity. A more refined question is, ``How fast can the gap to capacity decrease as a function of the blocklength?''. A key result is that, for any rate-$R$ code achieving a block error rate of $\eta<1$ on a non-trivial discrete memoryless channel with capacity $C$, the blocklength $N$ must satisfy $C-R\geq A/\sqrt{N}$ for some $A>0$ that depends only on $\delta$ and the channel~\cite{Strassen-zfw64,Laneman-ita06,Hayashi-it09,Polyanskiy-it10}. Thus, the gap to capacity cannot vanish faster than $O(N^{-1/2})$. Random codes are known to achieve this scaling.

Polar codes are the first codes, with low-complexity encoding and decoding algorithms, that were proven to achieve capacity on binary-input memoryless channels~\cite{Arikan-isit08,Arikan-it09}. Since then, the rate of polarization and the relationship between the blocklength and the error rate has received significant attention~\cite{Arikan-isit09,Hassani-it13,Goldin-it14,Hassani-it14,Fazeli-aller14,Guruswami-arxiv14,Presman-it15,Guruswami-it15,Goldin-arxiv15,Mondelli-arxiv15}. This relationship is typically studied in two distinct regimes by asking two different questions. First, for a fixed rate $R<C$, how fast does the error rate decay with the blocklength? Second, for a fixed probability of decoding failure $\eta\in(0,1)$, how fast can the rate approach the capacity?

The majority of prior work in this area focuses on binary polar codes with $2\times2$ kernels and, for these codes, the gap to capacity cannot decrease faster than $O(N^{-0.276})$~\cite{Hassani-it14,Mondelli-arxiv15}. For $2\times2$ kernels with larger alphabets, the analysis is much more difficult and the provable scaling rates are even smaller~\cite{Guruswami-arxiv14,Goldin-arxiv15}. Recently, $8\times8$ and $16\times16$ binary kernels have been constructed that achieve scaling rates of $O(N^{-0.279})$ and $O(N^{-0.298})$~\cite{Fazeli-aller14}. Until now, no reported results provably established scaling rates faster than $O(N^{-0.30})$. 

In the first part of this work, we consider the $q$-ary polar codes introduced by Mori and Tanaka based on $q\times q$ Reed-Solomon (RS) polarization kernels with elements from the Galois field $\mathbb{F}_{q}$~\cite{Mori-itw10,Mori-it14}. Thus, in all statements, $q$ is implicitly assumed to be a prime power. These codes have length $N=q^{n}$, where $n$ is the number of steps in the polarization process. We consider transmission over the $q$-ary erasure channel (QEC) with erasure probability $\epsilon$. Mori and Tanaka have also shown that these polar codes achieve capacity on symmetric $q$-ary channels~\cite{Mori-it14}. 

By analyzing the polarization process for the QEC, we show that, for any $\gamma>0$ and $\delta>0$, there is a $q_{0}$ such that, for all $q\geq q_{0}$, the fraction of effective channels with erasure rate at most $N^{-\gamma}$ is at least $1-\epsilon-O(N^{-1/2+\delta})$. Thus, the gap to capacity scales at a nearly-optimal rate. While our proof relies on large alphabet QECs with large polarization kernels, we believe a similar result may also hold for small alphabets (e.g., binary) with large polarization kernels.

Like binary polar codes, the performance of $q$-ary polar codes can be analyzed by tracking the evolution of the effective channels through the polarization process~\cite{Arikan-it09}. At each step, a single effective channel with erasure rate $x$ splits into $q$ new channels. For their codes, Mori and Tanaka showed that the $i$-th new effective channel, for $i\in\mathcal{Q}\triangleq\left\{ 0,1,\ldots,q-1\right\} $, is a $q$-ary erasure channel with erasure probability
\begin{equation}
\psi_{i}(x)=\sum_{j=i+1}^{q}\binom{q}{j}x^{j}(1-x)^{q-j},\label{eq:mori_qec_de}
\end{equation}
Applying this formula recursively, one can compute the erasure rates of the $N=q^{n}$ effective channels after $n$ steps. For a polar code with $k$ information symbols, the next step in the design process consists of choosing the $k$ effective channels with the smallest erasure rates. For $q=2$, these steps are identical to the original polar code construction in~\cite{Arikan-it09} and the resulting codes are closely related to binary Reed-Muller codes. For larger $q$, the resulting codes are closely related to $q$-ary Reed-Muller codes~\cite{Kschischang-isit93}.

In Section~\ref{sec:fixed_alphabet}, a similar analysis is also considered for $q$-ary polar codes with $m\times m$ polarizing matrices. This separates the effect of the alphabet size from the effect of the matrix size. If the polarizing matrix at each stage is drawn independently and uniformly from the set of invertible $m\times m$ matrices, then the linear operator associated with the Lyapunov function analysis can be written in closed form. To prove near-optimal scaling for polar codes with fixed $q$ as $m$ increases, however, two technical obstacles remain. Thus, we conclude by stating two concrete mathematical conjectures that, if proven, would imply near-optimal scaling for fixed $q$.

\section{The Polarization Process}

Let the random variable $X_{n}$ denote the channel erasure probability for a randomly chosen effective channel after $n$ levels of polarization. The sequence $X_{n}$, for $n=0,1,\ldots$, is a homogeneous Markov chain on the compact state space $\mathcal{X}=[0,1]$ with transition probability
\begin{align*}
\Pr\Big( & X_{n}\,=x_{n}\,\Big|\,(X_{0},\ldots,X_{n-1})=(x_{0},\ldots,x_{n-1})\Big)\\
 & =\Pr\left(X_{n}=x_{n}\midb |X_{n-1}=x_{n-1}\right)\\
 & \quad=\frac{1}{q}\left|\left\{ i\in\mathcal{Q}\,|\,x_{n}=\psi_{i}(x_{n-1})\right\} \right|.
\end{align*}
We note that $0$ and $1$ are both absorbing states of this Markov chain and we are interested in the convergence rate to these states~\cite{Arikan-it09}.

Let $C(\mathcal{X})$ denote the set of bounded continuous functions mapping $\mathcal{X}$ to $\mathbb{R}$. One can analyze this Markov chain by focusing on the sequence of functions, $g_{n}(x)\triangleq\expt[g_{0}(X_{n})|X_{0}=x]$, generated by $g_{0}\in C(\mathcal{X})$~\cite{Hassani-it14,Mondelli-arxiv15}. Since the Markov chain is homogeneous, this sequence satisfies the recursion

\begin{align*}
g_{n} & (x)\triangleq\expt\left[g_{0}(X_{n})\midb |X_{0}=x\right]\\
 & =\sum_{i=0}^{q-1}\expt\left[g(X_{n})\midb |X_{1}\!=\!\psi_{i}(x)\right]\Pr\left(X_{1}\!=\!\psi_{i}(x)\midb |X_{0}\!=\!x\right)\\
 & =\sum_{i=0}^{q-1}g_{n-1}\left(\psi_{i}(x)\right)\frac{1}{q}.
\end{align*}
The one-step update is given by the linear operator $T_{q}:C(\mathcal{X})\to C(\mathcal{X})$, which is defined by
\begin{equation}
(T_{q}g_{n-1})(x)\triangleq\frac{1}{q}\sum_{i=0}^{q-1}g_{n-1}\left(\psi_{i}(x)\right).\label{eq:mori_qec}
\end{equation}
Since the polarization process preserves the average mutual information, it also preserves average erasure rate. This implies that the function $g_{0}(x)=x$ should be an eigenfunction of $T_{q}$ (with eigenvalue $1$) and, using \eqref{eq:mori_qec}, one can verify that it is. We note that this is a straightforward generalization of the approach used for binary polar codes~\cite{Hassani-it14,Mondelli-arxiv15}. 

The rate of polarization is determined by the fraction of channels whose erasure rates are not extremal. The following lemma connects the fraction of non-extremal channels (as a function of $n$) with an easily computable constant associated with $T_{q}$. This can be seen as a standard convergence analysis based on Lyapunov functions and it was first applied to polar codes in~\cite{Hassani-it14}.
\begin{lem}
\label{lem:lyapunov} Suppose there exists a non-negative continuous function $V:\mathcal{X}\to\mathbb{R}_{\geq0}$ and a constant $\lambda\in(0,1)$ such that 
\begin{equation}
(T_{q}V)(x)\leq\lambda V(x)\label{eq:exp_drift}
\end{equation}
for all $x\in\mathcal{X}$. Then, for $S(\alpha)\triangleq\left\{ x\in\mathcal{X}\,|\,V(x)\geq\alpha\right\} ,$ it follows that 
\[
\Pr\left(X_{n}\in S(\alpha)\midb |X_{0}=x\right)\leq\frac{\lambda^{n}V(x)}{\alpha}.
\]
Further, if $S(\alpha)$ is a closed interval, then 
\[
\Pr\left(X_{n}\geq\min S(\alpha)\midb |X_{0}=x\right)\leq\frac{\lambda^{n}V(x)}{\alpha}+\frac{x}{\max S(\alpha)}.
\]
\end{lem}
\begin{IEEEproof}
To see this, we choose $g_{0}(x)=V(x)$ and observe that~\eqref{eq:exp_drift} implies $\expt\left[V(X_{n})\,|\,X_{0}=x\right]=g_{n}(x)\leq\lambda^{n}V(x)$ for all $x\in\mathcal{X}$. From this, we get 
\begin{align*}
\Pr\left(X_{n}\in S(\alpha)\midb |X_{0}=x\right) & =\Pr\left(V(X_{n})\geq\alpha\midb |X_{0}=x\right)\\
 & \leq\frac{\expt\left[V(X_{n})\midb |X_{0}=x\right]}{\alpha}\\
 & \leq\frac{\lambda^{n}V(x)}{\alpha}.
\end{align*}
Since the polarization process preserves the average mutual information, we have $\expt[X_{n}\,|\,X_{0}=x]=x$. For the second part, we combine this with the Markov inequality to see that
\begin{align*}
\Pr\left(X_{n}>\max S(\alpha)\midb |X_{0}=x\right) & \leq\frac{\expt\left[X_{n}\midb |X_{0}=x\right]}{\max S(\alpha)}\\
 & =\frac{x}{\max S(\alpha)}.
\end{align*}
Since $S(\alpha)$ is a closed interval, it follows that
\begin{align*}
\Pr & \left(X_{n}\geq\min S(\alpha)\midb |X_{0}\!=\!x\right)=\Pr\left(X_{n}\!\in\!S(\alpha)\midb |X_{0}\!=\!x\right)\\
 & \quad+\Pr\left(X_{n}>\max S(\alpha)\midb |X_{0}=x\right)\\
 & \quad\quad\leq\frac{\lambda^{n}V(x)}{\alpha}+\frac{x}{\max S(\alpha)}.
\end{align*}
This completes the proof.
\end{IEEEproof}
\begin{rem}
For the considered problem, this lemma is a slight variation of what is used in~\cite{Hassani-it13,Mondelli-arxiv15}. We use this form to show the close connection to Lyapunov functions. From that perspective, the function $V(x)$ can be seen as a Lyapunov function showing convergence to stationary distributions supported on the set $\left\{ x\in\mathcal{X}\,|\,V(x)=0\right\} $~\cite{Hairer-ssa11}.
\end{rem}
\begin{defn}
\label{def:lambda_q_b} Let $V(x)=(x(1-x))^{\beta}$ for $\beta>0$ and define 
\[
\lambda_{q,\beta}\triangleq\sup_{x\in(0,1)}\frac{(T_{q}V)(x)}{V(x)}.
\]
Then, $\lambda_{q,\beta}$ is the largest $\lambda\in\mathbb{R}$ such that $(T_{q}V)(x)\leq\lambda V(x)$ for all $x\in(0,1)$. We also note that $V(x)\leq V(\frac{1}{2})=(\frac{1}{4})^{\beta}$ for $x\in[0,1].$
\end{defn}
\begin{lem}
\label{lem:lambda_q_b_bound} The quantity $\lambda_{q,\beta}$ for $\beta\in(0,\frac{1}{2}]$ satisfies
\[
\lambda_{q,\beta}\leq\frac{6}{\sqrt{q\beta}}\left(\frac{1}{4}\right)^{\frac{1}{2}-\beta}.
\]
\end{lem}
\begin{IEEEproof}
See Section~\ref{subsec:proof_qary_main}.
\end{IEEEproof}
\begin{cor}
\label{cor:lyapunov} If the conditions of Lemma~\ref{lem:lyapunov} hold for $V(x)=(x(1-x))^{\beta}$ with $\beta>0$, then 
\[
\Pr\left(X_{n}\in[\eta,1-\eta]\midb |X_{0}=x\right)\leq\frac{\lambda^{n}V(x)}{V(\eta)}
\]
for $\eta\in(0,\frac{1}{2})$. This also implies
\[
\Pr\left(X_{n}\geq\eta\midb |X_{0}=x\right)\leq\frac{\lambda^{n}V(x)}{V(\eta)}+\frac{x}{1-\eta}.
\]
\end{cor}
\begin{IEEEproof}
The first statement follows from applying Lemma~\ref{lem:lyapunov} with $\alpha=V(\eta)$. For the second statement, we observe that $S(\alpha)$ is a closed interval because $V(x)$ is a concave function. Also, $\max S(\alpha)=1-\eta$ because $V(\eta)=V(1-\eta)$ implies that $1-\eta\in S(\alpha)$ and $V(x)<V(\eta)$ for $x>1-\eta$. This completes the proof.
\end{IEEEproof}
The primary purpose of this paper is the statement and proof of the following theorem.
\begin{thm}
\label{thm:main_qary} For the $q$-ary polar codes defined in \cite{Mori-itw10,Mori-it14}, let $X_{n}$ be the erasure rate of a randomly chosen effective channel after $n$ steps of polarization. For any $\gamma>0$, $\beta\in(0,\frac{1}{2}]$, and $N^{-\gamma}\leq\frac{3}{4}$, one finds that
\begin{align*}
\Pr\Big(X_{n}\in[N^{-\gamma} & ,1-N^{-\gamma}]\,\Big|\,X_{0}=x\Big)\\
 & \leq N^{\gamma\beta-\frac{1}{2}+\frac{\ln6-\frac{1}{2}\ln\beta+\left(\beta-\frac{1}{2}\right)\ln4}{\ln q}}
\end{align*}
and 
\begin{align*}
\Pr\Big(X_{n}\geq & N^{-\gamma}\,\Big|\,X_{0}=x\Big)\\
 & \leq N^{\gamma\beta-\frac{1}{2}+\frac{\ln6-\frac{1}{2}\ln\beta+\left(\beta-\frac{1}{2}\right)\ln4}{\ln q}}+\frac{x}{1-N^{-\gamma}}.
\end{align*}
\end{thm}
\begin{IEEEproof}
Combining Lemma~\ref{lem:lambda_q_b_bound} and Corollary~\ref{cor:lyapunov} with $\eta=N^{-\gamma}$, one gets the prediction
\begin{align*}
\Pr\Big(X_{n} & \in[N^{-\gamma},1-N^{-\gamma}]\,\Big|\,X_{0}=x\Big)\\
 & \leq\frac{\left(x(1-x)\right)^{\beta}}{\left(N^{-\gamma}(1-N^{-\gamma})\right)^{\beta}}\left(6\sqrt{\frac{1}{q\beta}}\left(\frac{1}{4}\right)^{\frac{1}{2}-\beta}\right)^{n}\\
 & \leq\left(\frac{\frac{1}{4}}{1-N^{-\gamma}}\right)^{\beta}N^{\gamma\beta}q{}^{n\frac{\ln6-\frac{1}{2}\ln q-\frac{1}{2}\ln\beta+\left(\beta-\frac{1}{2}\right)\ln4}{\ln q}}\\
 & \leq N^{\gamma\beta-\frac{1}{2}+\frac{\ln6-\frac{1}{2}\ln\beta+\left(\beta-\frac{1}{2}\right)\ln4}{\ln q}},
\end{align*}
for $N^{-\gamma}\leq\frac{3}{4}$. The second statement follows directly from the second part of Corollary~\ref{cor:lyapunov}.
\end{IEEEproof}
\begin{cor}
Consider the $q$-ary polar codes defined in \cite{Mori-itw10,Mori-it14} on a QEC with erasure probability $\epsilon$. For any $\gamma>0$ and $\delta>0$, there is a $\beta\in(0,\frac{1}{2}]$ and a $q_{0}$ such that, for all $q\geq q_{0}$, the fraction of effective channels with erasure rate at most $N^{-\gamma}$ is at least $1-\epsilon-O(N^{-1/2+\delta})$.
\end{cor}
\begin{IEEEproof}
Since the stated condition becomes weaker as $\gamma$ decreases and $\delta$ increases, we assume without loss of generality that $\gamma\geq\frac{1}{2}$ and $\delta\leq\frac{1}{2}$. Using this, we choose $\beta=\frac{\delta}{2\gamma}$ and observe that $\beta\in(0,\frac{1}{2}]$. At error rate $N^{-\gamma}$, the gap to capacity is given by
\begin{align*}
(1-\epsilon) & -\Pr\left(X_{n}<N^{-\gamma}\,|\,X_{0}=\epsilon\right)\\
 & =\Pr\left(X_{n}\geq N^{-\gamma}\,|\,X_{0}=\epsilon\right)-\epsilon.
\end{align*}
Since $2^{-1/2}\leq\frac{3}{4}$, we can applying Theorem~\ref{thm:main_qary} for $N\geq2$ to see that
\begin{align*}
\Pr & \left(X_{n}\geq N^{-\gamma}\,\Big|\,X_{0}=\epsilon\right)-\epsilon\\
 & \stackrel{(a)}{\leq}N^{\frac{\delta}{2}-\frac{1}{2}+\frac{\ln6-\frac{1}{2}\ln\beta+\left(\beta-\frac{1}{2}\right)\ln4}{\ln q}}+\frac{\epsilon N^{-\gamma}}{1-N^{-\gamma}}\\
 & \quad\quad\stackrel{(b)}{\leq}N^{\frac{\delta}{2}-\frac{1}{2}+\frac{\delta}{2}}+4\epsilon N^{-\frac{1}{2}}\\
 & \quad\quad\quad\quad\leq5N^{-\frac{1}{2}+\delta},
\end{align*}
where $(a)$ follows from $\frac{\epsilon}{1-N^{-\gamma}}-\epsilon=\frac{\epsilon N^{-\gamma}}{1-N^{-\gamma}}$ and $(b)$ follows from $N^{-\gamma}\leq2^{-1/2}\leq\frac{3}{4}$ and choosing $q\geq q_{0}$ with $\ln q_{0}\triangleq\frac{1}{\delta}(2\beta\ln4-\ln\beta+2\ln6-\ln4)$. This completes the proof.
\end{IEEEproof}

\subsection{Numerical Examples}

In this section, we present some applications of Corollary~\ref{cor:lyapunov} based on numerical computation of $\lambda$.

\begin{figure}[!t]
\begin{center}
\scalebox{1.2}{\includegraphics{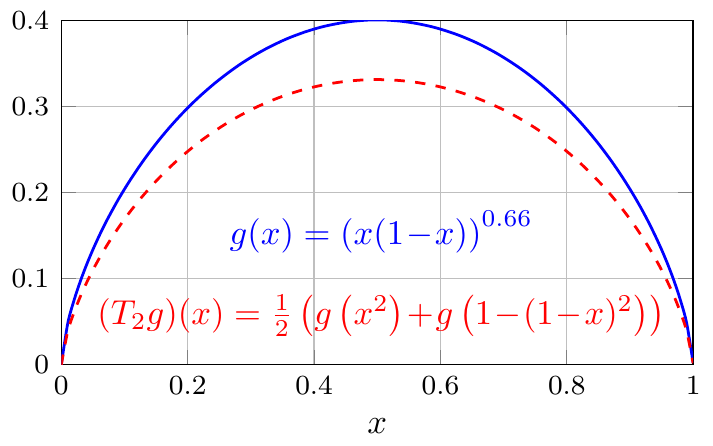}}
\end{center}\vspace{-4mm}

\caption{\label{fig:lambda_2}Numerical evaluation of the constant $\lambda_{2,0.66}$. }
\end{figure}

\begin{example}
Consider the case of $q=2$ where $T_{q}$ is defined by 
\[
(T_{2}g)(x)=\frac{g(x^{2})+g(2x-x^{2})}{2}.
\]
Using $V(x)=\left(x(1-x)\right)^{0.66}$, one can verify numerically that $(T_{2}V)(x)\leq0.832V(x)$ for $x\in[0,1]$. For example, see Figure~\ref{fig:lambda_2}. We note that this calculation was described first in~\cite{Hassani-it14}. Therefore, 
\begin{align*}
\Pr\Big(X_{n}\in[0.01,0.99] & \Big|X_{0}=x\Big)\leq\frac{(1/4)^{0.66}}{(0.0099)^{0.66}}0.832^{n}\\
 & =9\cdot2^{n\frac{\ln0.832}{\ln2}}\\
 & \leq9N^{-0.265}.
\end{align*}
Let $V_{5}(x)$ be the result of applying $T_{2}$ five times to the function $\left(x(1-x)\right)^{0.66}$ (i.e., $V_{5}(x)=\left(T_{2}^{5}\left(x(1-x)\right)^{0.66}\right)(x)$). Then, one can verify numerically that $(T_{2}V_{5})(x)\leq0.8271V_{5}(x)$ and this gives a decay rate of $O(N^{-0.273})$. 
\end{example}

\begin{example}
Consider the case of $q=4$ where $T_{q}$ is defined by 
\begin{align*}
(T_{4}g)(x) & =\frac{1}{4}\bigg(g(x^{4})+g(4x^{3}(1-x)+x^{4})+\\
g(1- & 4x(1-x)^{3}-(1-x)^{4})+g(1-(1-x)^{4})\bigg)
\end{align*}
Using $V(x)=\left(x(1-x)\right)^{0.64}$, one can verify numerically that $(T_{4}V)(x)\leq0.657V(x)$ for $x\in[0,1]$. Therefore, 
\begin{align*}
\Pr\Big(X_{n}\in[0.01,0.99] & \Big|X_{0}=x\Big)\leq\frac{(1/4)^{0.64}}{(0.0099)^{0.64}}0.657^{n}\\
 & =8\cdot4^{n\frac{\ln0.657}{\ln4}}\\
 & \leq8N^{-0.303}.
\end{align*}
\end{example}

\begin{figure}[!t]
\vspace{2mm}
\begin{center}
\scalebox{1.2}{\includegraphics{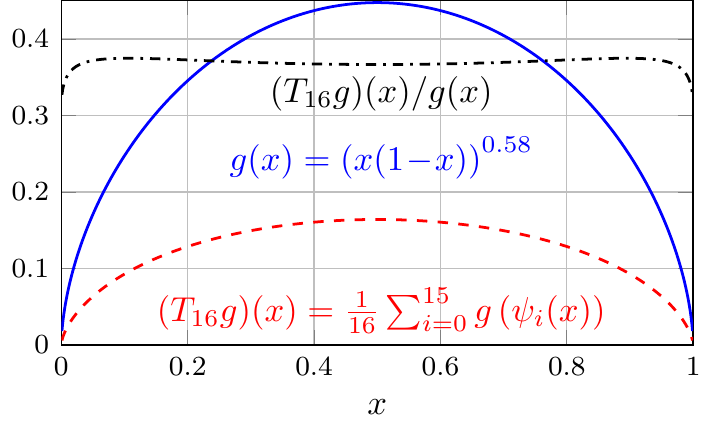}}
\end{center}\vspace{-4mm}

\caption{\label{fig:lambda_16}Numerical evaluation of the constant $\lambda_{16,0.58}$.}
\end{figure}

\begin{example}
Consider the case of $q=16$ where $T_{q}$ is defined by \eqref{eq:mori_qec}. Using $V(x)=\left(x(1-x)\right)^{0.58}$, one can verify numerically that $(T_{16}V)(x)\leq0.375V(x)$ for $x\in[0,1]$. For example, see Figure~\ref{fig:lambda_16}. Therefore, 
\begin{align*}
\Pr\Big(X_{n}\in[0.01,0.99] & \Big|X_{0}=x\Big)\leq\frac{(1/4)^{0.58}}{(0.0099)^{0.58}}0.375^{n}\\
 & =7\cdot16^{n\frac{\ln0.375}{\ln16}}\\
 & \leq7N^{-0.353}.
\end{align*}
\end{example}

\begin{example}
\label{exa:sqrt_numerical_largeq} Consider the case where $T_{q}$ is defined by \eqref{eq:mori_qec} for $q=2,3,\ldots,1024$. Using $V(x)=\sqrt{x(1-x)}$, one can compute numerically the smallest $\lambda_{q}$ such that $(T_{q}V)(x)\leq\lambda_{q}V(x)$ for $x\in[0,1]$. This computation results in $\lambda_{q}=(T_{q}V)(\frac{1}{2})/V(\frac{1}{2})$ and one observes that $\sqrt{q}\lambda_{q}$ is increasing in $q$ and upper bounded by $1.6142$. Assuming this is true, we observe that
\begin{align*}
\Pr\Big(X_{n}\!\in\![\eta,1-\eta]\Big| & X_{0}=x\Big)\!\leq\!\frac{\sqrt{1/4}}{\sqrt{\eta(1-\eta)}}\left(\frac{1.6142}{\sqrt{q}}\right)^{\!n}\\
 & =\frac{1}{\sqrt{4\eta(1-\eta)}}q{}^{n\frac{\ln1.6142-\frac{1}{2}\ln q}{\ln q}}\\
 & \leq\frac{1}{\sqrt{4\eta(1-\eta)}}N^{-\frac{1}{2}(1-\frac{1}{\ln q})}.
\end{align*}
\end{example}

\begin{example}
Let $V(x)=(x(1-x))^{1/12}$ and consider the case where $T_{q}$ is defined by \eqref{eq:mori_qec} for $q=2,3,\ldots,1024$. Again, one can compute numerically the smallest $\lambda_{q}=\lambda_{q,1/12}$ such that $(T_{q}V)(x)\leq\lambda_{q}V(x)$ for $x\in[0,1]$. This computation results in $\lambda_{q}=(T_{q}V)(\frac{1}{2})/V(\frac{1}{2})$ and one observes that $\sqrt{q}\lambda_{q}$ is increasing in $q$ and upper bounded by $4.1218$. Assuming this is true, we observe that, for $N\geq2$, we have
\begin{align*}
\Pr\Big(X_{n}\in & [N^{-2},1-N^{-2}]\Big|X_{0}=x\Big)\\
 & \leq\frac{(1/4)^{1/12}}{\left(N^{-2}(1-N^{-2})\right)^{1/12}}\left(\frac{4.1218}{\sqrt{q}}\right)^{n}\\
 & =\left(\frac{1/4}{1-N^{-2}}\right)^{1/12}N^{\frac{1}{6}}q{}^{n\frac{\ln4.1218-\frac{1}{2}\ln q}{\ln q}}\\
 & \leq\frac{1}{3}N^{-\frac{1}{3}(1-\frac{4.5}{\ln q})}.
\end{align*}
\end{example}

\section{Large-Alphabet Erasure Channels}

\subsection{Intuitive Approach}

Before delving into the proof of Theorem~\ref{thm:main_qary}, we present an intuitive (but non-rigorous) argument that leads us in the right direction. Consider $q$ random trials with success probability $x$ and let the random variable $\mbox{Bin}(q,x)$ denote number of successes. Then, one finds that
\[
\Pr\left(\mathrm{Bin}(q,x)=i\right)=\binom{q}{i}x^{i}(1-x)^{n-i}.
\]
The key is to replace the binomial random variable, $\mathrm{Bin}(q,x)$, by a Gaussian random variable with the same mean and variance. While this step is motivated by the central limit theorem, it is not rigorous (even as $q\to\infty$) because the approximation does not hold uniformly for all $x\in[0,1]$. Based on this assumption, we approximate $\Pr\left(\mathrm{Bin}(q,x)\geq i+1\right)$ by 
\[
\psi_{i}(x)\approx Q\left(\frac{i+1-qx}{\sqrt{qx(1-x)}}\right),
\]
where $Q(x)\triangleq(2\pi)^{-1/2}\int_{x}^{\infty}e^{-t^{2}/2}dt$. Let $V(x)=(x(1-x))^{\beta}$ for $\beta\in(0,\frac{1}{2}]$. Using a sequence of approximations, one finds that 
\begin{align*}
(T_{q}V)(x) & =\frac{1}{q}\sum_{i=0}^{q-1}V\left(\psi_{i}(x)\right)\\
 & \approx\frac{1}{q}\sum_{i=0}^{q-1}V\left(Q\left(\frac{i+1-qx}{\sqrt{qx(1-x)}}\right)\right)\\
 & \approx\int_{0}^{1}V\left(Q\left(\frac{q(y-x)}{\sqrt{qx(1-x)}}\right)\right)\dd y\\
 & =\sqrt{\frac{x(1-x)}{q}}\int_{-x\sqrt{qx/(1-x)}}^{\sqrt{q(1-x)/x}}V\left(Q\left(z\right)\right)\dd z\\
 & \approx\sqrt{\frac{x(1-x)}{q}}\int_{-\infty}^{\infty}V\left(Q\left(z\right)\right)\dd z\\
 & =\sqrt{\frac{x(1-x)}{q}}\mathfrak{m}(\beta),
\end{align*}
where $\mathfrak{m}(\beta)\triangleq\int_{-\infty}^{\infty}\left(Q(z)Q(-z)\right)^{\beta}dz$.

For $\beta=\frac{1}{2}$, this implies that $V(x)=\sqrt{x(1-x)}$ is an approximate eigenfunction of $T_{q}$ associated with eigenvalue 
\[
\tilde{\lambda}_{q}=\frac{\mathfrak{m}(\frac{1}{2})}{\sqrt{q}},
\]
where $\mathfrak{m}(\frac{1}{2})\approx1.6147<e^{1/2}$. Based on this estimate, one could estimate that rate of polarization scales like 
\begin{align*}
\tilde{\lambda}_{q}^{n} & \approx\left(\frac{\mathfrak{m}(\frac{1}{2})}{\sqrt{q}}\right)^{n}\leq\left(\frac{e}{q}\right)^{n/2}\\
 & =q^{(n/2)(1-\ln q)/\ln q}=N^{-\frac{1}{2}(1-\frac{1}{\ln q})}.
\end{align*}
In fact, the numerical results in Example~\ref{exa:sqrt_numerical_largeq} support this conclusion and suggest that the true $\lambda_{q,1/2}$ satisfies $\lambda_{q,1/2}\sqrt{q}\nearrow\mathfrak{m}(\frac{1}{2})$. Thus, we believe that this non-rigorous analysis produces an exact and tight characterization as $q\to\infty$.

For $\beta\in(0,\frac{1}{2})$, $V(x)$ is not an approximate eigenfunction but one can still estimate the decay rate
\begin{align*}
\tilde{\lambda}_{q,\beta} & =\max_{x\in[0,1]}\left(x(1-x)\right)^{\frac{1}{2}-\beta}\frac{\mathfrak{m}(\beta)}{\sqrt{q}}\\
 & =\frac{\mathfrak{m}(\beta)}{\sqrt{q}}\left(\frac{1}{4}\right)^{\frac{1}{2}-\beta}.
\end{align*}
Combining this estimate with Corollary~\ref{cor:lyapunov} gives the non-rigorous prediction 
\begin{align*}
\Pr\Big(X_{n}\in & [N^{-\gamma},1-N^{-\gamma}]\,\Big|\,X_{0}=x\Big)\\
 & \leq\frac{\left(x(1-x)\right)^{\beta}}{\left(N^{-\gamma}(1-N^{-\gamma})\right)^{\beta}}\left(\frac{\mathfrak{m}(\beta)}{\sqrt{q}}\left(\frac{1}{4}\right)^{\frac{1}{2}-\beta}\right)^{n}\\
 & \leq\left(\frac{\frac{1}{4}}{1-N^{-\gamma}}\right)^{\beta}N^{\gamma\beta}q{}^{n\frac{\ln\left(\mathfrak{m}(\beta)\right)-\frac{1}{2}\ln q+\left(\beta-\frac{1}{2}\right)\ln4}{\ln q}}\\
 & \leq N^{\gamma\beta-\frac{1}{2}+\frac{\ln\left(\mathfrak{m}(\beta)\right)+\left(\beta-\frac{1}{2}\right)\ln4}{\ln q}},
\end{align*}
for $N^{-\gamma}\leq\frac{3}{4}$. The key point here is that, for any $\gamma>0$ and $\delta>0$, there is a $\beta\in(0,\frac{1}{2}]$ and a large enough $q$ such that this decay rate is $O(N^{-1/2+\delta})$.

\subsection{Rigorous Approach}

Unfortunately, the intuitive argument does not lead directly to a rigorous statement because the central limit theorem is tight only for small deviations. To make things precise, one must instead use the Chernoff-Hoeffding bound for the binomial tail probability. 

We start by establishing some basic properties of the functions under consideration. Looking at~\eqref{eq:mori_qec_de}, one observes that
\begin{align*}
\psi_{i}(x) & =\Pr\left(\mathrm{Bin}(q,x)\geq i+1\right)\\
 & =\Pr\left(\mathrm{Bin}(q,1-x)\leq q-i-1\right)\\
 & =1-\Pr\left(\mathrm{Bin}(q,1-x)\geq q-i\right)\\
 & =1-\psi_{q-i-1}(1-x).
\end{align*}
This implies the following lemma.
\begin{lem}
\label{lem:Tq_g_symmetry} If $g(x)=g(1-x)$, then $(T_{q}g)(x)=(T_{q}g)(1-x)$.
\end{lem}
\begin{IEEEproof}
Working directly, one finds that
\begin{align*}
(T_{q}g)(x) & =\frac{1}{q}\sum_{i=0}^{q-1}g\left(\psi_{i}(x)\right)\\
 & =\frac{1}{q}\sum_{i=0}^{q-1}g\left(1-\psi_{i}(x)\right)\\
 & =\frac{1}{q}\sum_{i=0}^{q-1}g\left(\psi_{q-i-1}(1-x)\right)\\
 & =\frac{1}{q}\sum_{i=0}^{q-1}g\left(\psi_{i}(1-x)\right)\\
 & =(T_{q}g)(1-x).
\end{align*}
\end{IEEEproof}
The well-known Chernoff bound for the binomial tail probability implies that, for $i+1\geq qx$, one has 
\begin{equation}
\psi_{i}(x)=\Pr\left(\mathrm{Bin}(q,x)\geq i+1\right)\leq e^{-qD(\frac{i+1}{q}||x)},\label{eq:Binomial_KL1}
\end{equation}
where $D(y||x)\triangleq y\ln\frac{y}{x}+(1-y)\ln\frac{1-y}{1-x}$ is the Kullback-Leibler divergence between two Bernoulli distributions. Similarly, for $i\leq qx$, one has 
\begin{equation}
1-\psi_{i}(x)=\Pr\left(\mathrm{Bin}(q,x)\leq i\right)\leq e^{-qD(\frac{i}{q}||x)}.\label{eq:Binomial_KL2}
\end{equation}

\begin{lem}
\label{lem:Simple_KL_bin_bound} For $x\leq y$, we have
\[
\Pr\left(\mathrm{Bin}(q,x)\geq qy\right)\leq e^{-qd(y,x)},
\]
where $d(y,x)\triangleq\frac{1}{2}(y-x)^{2}/(x(1-x)+(1-2x)(y-x)/3)$. Similarly, for $x\geq y$, we have $\Pr\left(\mathrm{Bin}(q,x)\leq qy\right)\leq e^{-qd(y,x)}$.
\end{lem}
\begin{IEEEproof}
It is well known from the Chernoff bound that $\Pr\left(\mathrm{Bin}(q,x)\geq qy\right)\leq e^{-qD(y||x)}$ for $x\leq y$, where $D(y||x)\triangleq y\ln\frac{y}{x}+(1-y)\ln\frac{1-y}{1-x}$ is the Kullback-Leibler divergence. Thus, the first result holds if $d(y,x)\leq D(y||x)$ for $x\leq y$. Since $D(1-y||1-x)=D(y||x)$ and $d(1-y,1-x)=d(y,x)$, the second result follows from the first by symmetry. Thus, it suffices to prove that $d(y,x)\leq D(y||x)$ for $x\leq y$. To do this, we first observe that
\begin{align*}
\frac{\dd}{\dd x} & \left(d(y,x)-D(y||x)\right)\\
 & =\frac{(1-x(1-x))(y-x)^{3}}{x(1-x)(y-x(x+2y-2))^{2}}\geq0
\end{align*}
for $x\leq y$. Next, we observe that
\begin{align*}
\int_{x}^{y} & \left(\frac{\dd}{\dd x'}\left(d(y,x')-D(y||x')\right)\right)\dd x\\
 & =\left(d(y,y)-D(y||y)\right)-\left(d(y,x)-D(y||x)\right)\geq0
\end{align*}
because $x\leq y$ throughout the range of integration. Since $d(y,y)=D(y||y)=0$, this implies $d(y,x)\leq D(y||x)$ for $x\leq y$.
\end{IEEEproof}
\begin{lem}
\label{lem:KL_Poisson} For $x,y\in[0,1]$, we have
\[
D(y||x)\geq(y-x)+(1-y)\ln\frac{1-y}{1-x}
\]
and, for $z\in[0,1]$, we have
\[
1-z+z\ln z\geq\frac{1}{2}(1-z)^{2}.
\]
\end{lem}
\begin{IEEEproof}
The first bound follows from lower bounding the $y\ln\frac{y}{x}$ term in $D(y||x)$ by 
\[
y\ln\frac{y}{x}=-y\ln\frac{y-(y-x)}{y}\geq-y\left(-\frac{y-x}{y}\right)=y-x.
\]
Let $f(z)=z+(1-z)\ln(1-z)$ and observe that $f(1-z)=1-z+z\ln z$. Since $f'(0)=f(0)=0$ and $f''(z)=\frac{1}{1-z}\geq1$ for $z\in[0,1]$, it follows that
\[
f(z)=\int_{0}^{z}\int_{0}^{y}f''(x)\dd x\,\dd y\geq\frac{1}{2}z^{2}.
\]
Thus, $f(1-z)=1-z+z\ln z\geq\frac{1}{2}(1-z)^{2}.$
\end{IEEEproof}
\begin{lem}
\label{lem:g_psi_middle_term} For $\beta\in(0,\frac{1}{2}]$, $V(x)=(x(1-x))^{\beta}$, and $x\in[\frac{1}{2},1]$, we have

\[
\frac{1}{q}V\left(\psi_{\left\lceil qx\right\rceil -1}(x)\right)\leq\frac{\left(2x(1-x)\right)^{\beta}}{\sqrt{2q}}.
\]
\end{lem}
\begin{IEEEproof}
If $x\in[\frac{1}{2},1-\frac{1}{q}]$, then we have 
\begin{align*}
\frac{1}{q}V & \left(\psi_{\left\lceil qx\right\rceil -1}(x)\right)\stackrel{(a)}{\leq}\frac{1}{q}\left(\frac{1}{4}\right)^{\beta}\\
 & =\frac{\left(2x(1-x)\right)^{\beta}}{\sqrt{2q}}\frac{\sqrt{2q}}{\left(2x(1-x)\right)^{\beta}}\frac{1}{q}\left(\frac{1}{4}\right)^{\beta}\\
 & =\frac{\left(2x(1-x)\right)^{\beta}}{\sqrt{2q}}\frac{\sqrt{2}}{\left(2xq(1-x)\right)^{\beta}}\frac{1}{q^{1/2-\beta}}\left(\frac{1}{4}\right)^{\beta}\\
 & \stackrel{(b)}{\leq}\frac{\left(2x(1-x)\right)^{\beta}}{\sqrt{2q}}\frac{\sqrt{2}}{q^{1/2-\beta}}\left(\frac{1}{4}\right)^{\beta}\\
 & \leq\frac{\left(2x(1-x)\right)^{\beta}}{\sqrt{2q}}\sup_{q\geq2}\frac{q^{\beta}\sqrt{2}}{q^{1/2}4^{\beta}}\\
 & \stackrel{(b)}{=}\frac{\left(x(1-x)\right)^{\beta}}{\sqrt{2q}},
\end{align*}
where $(a)$ holds because $V(z)\leq(\frac{1}{4})^{\beta}$, $(b)$ follows from $2x\geq1$ and $q(1-x)\geq1$, and $(c)$ holds because the argument of the supremum is decreasing in $q$. If $x\in(1-\frac{1}{q},1]$, then assume $x=1-\frac{\alpha}{q}$ for $\alpha\in[0,1)$ and observe that
\begin{align*}
\frac{1}{q}V & \left(\psi_{\left\lceil qx\right\rceil -1}(x)\right)=\frac{1}{q}V\left(\psi_{q-1}(x)\right)\\
 & =\frac{1}{q}(x^{q}(1-x^{q}))^{\beta}\\
 & =\frac{1}{q}\left(\left(1-\frac{\alpha}{q}\right)^{q}\left(1-\left(1-\frac{\alpha}{q}\right)^{q}\right)\right)^{\beta}\\
 & \stackrel{(a)}{\leq}\frac{1}{q}e^{-\alpha\beta}\alpha^{\beta}\\
 & =\frac{\left(\frac{\alpha}{q}(1-\frac{\alpha}{q})\right)^{\beta}}{\sqrt{q}}\frac{\sqrt{q}}{\left(\frac{\alpha}{q}(1-\frac{\alpha}{q})\right)^{\beta}}\frac{1}{q}e^{-\alpha\beta}\alpha^{\beta}\\
 & =\frac{\left(\frac{\alpha}{q}(1-\frac{\alpha}{q})\right)^{\beta}}{\sqrt{q}}\frac{q^{\beta-1/2}}{\left((1-\frac{\alpha}{q})\right)^{\beta}}e^{-\alpha\beta}\\
 & \leq\frac{\left(x(1-x)\right)^{\beta}}{\sqrt{q}}\sup_{\alpha\in[0,1)}\sup_{q\geq2}\frac{q^{\beta-1/2}}{\left((1-\frac{\alpha}{q})\right)^{\beta}}e^{-\alpha\beta}\\
 & \stackrel{(b)}{\leq}\frac{\left(x(1-x)\right)^{\beta}}{\sqrt{q}}\sup_{\alpha\in[0,1)}\frac{2^{\beta-1/2}}{\left((1-\frac{\alpha}{2})\right)^{\beta}}e^{-\alpha\beta}\\
 & \stackrel{(c)}{\leq}\frac{\left(x(1-x)\right)^{\beta}}{\sqrt{q}}2^{\beta-1/2},
\end{align*}
where $(a)$ follows from $1-\alpha\leq\left(1-\frac{\alpha}{q}\right)^{q}\leq e^{-\alpha}$, $(b)$ holds because the argument of the supremum is decreasing in $q$, and $(c)$ holds because the argument of the supremum is decreasing in $\alpha$.
\end{IEEEproof}

\subsection{\label{subsec:proof_qary_main} Proof of Lemma~\ref{lem:lambda_q_b_bound}}

Let $V(x)=(x(1-x))^{\beta}$ with $\beta\in(0,\frac{1}{2}]$. Based on Lemma~\ref{lem:Tq_g_symmetry}, it is sufficient to analyze $(T_{q}V)(x)$ for $x\geq1/2$. To do this, we will use the decomposition 
\begin{equation}
\begin{split}\frac{1}{q}\sum_{i=0}^{q-1} & V(\psi_{i}(x))=\frac{1}{q}\Bigg(\sum_{i=0}^{\left\lceil qx\right\rceil -2}V(\psi_{i}(x))\\
 & \quad+V(\psi_{\left\lceil qx\right\rceil -1}(x))+\sum_{i=\left\lceil qx\right\rceil }^{q-1}V(\psi_{i}(x))\Bigg).
\end{split}
\label{eq:TV_sum_decomp}
\end{equation}

First, we consider the upper sum in~\eqref{eq:TV_sum_decomp}. Applying~\eqref{eq:Binomial_KL1} to $\psi_{i}(x)$ shows that
\begin{align*}
\psi_{i}(x) & =\Pr\left(\mathrm{Bin}(q,x)\geq i+1\right)\\
 & \leq e^{-qD(\frac{i+1}{q}||x)}
\end{align*}
for $i+1\geq qx$. Thus, for $i\in\{\left\lceil qx\right\rceil ,\ldots,q-1\}$, we have $V(\psi_{i}(x))\leq(\psi_{i}(x))^{\beta}\leq e^{-q\beta D(\frac{i+1}{q}||x)}$ and
\begin{align}
\frac{1}{q}\sum_{i=\left\lceil qx\right\rceil }^{q-1}V\left(\psi_{i}(x)\right) & \leq\frac{1}{q}\sum_{i=\left\lceil qx\right\rceil }^{q-1}e^{-q\beta D(\frac{i+1}{q}||x)}\nonumber \\
 & \stackrel{(a)}{\leq}\frac{1}{q}\sum_{i=\left\lceil qx\right\rceil }^{q-1}\int_{0}^{1}e^{-q\beta D(\frac{i+z}{q}||x)}\dd z\nonumber \\
 & \stackrel{(b)}{=}\int_{\left\lceil qx\right\rceil /q}^{1}e^{-q\beta D(y||x)}\dd y,\label{eq:Tq_bound_upper}
\end{align}
where $e^{-q\beta D(\frac{i+1}{q}||x)}\leq\int_{0}^{1}e^{-q\beta D(\frac{i+z}{q}||x)}\dd z$ holds in $(a)$ because $e^{-q\beta D(\frac{i+z}{q}||x)}$ is decreasing in $z$ for $i\geq qx$. Also, $(b)$ follows from grouping terms into one integral and changing the variable of integration. Although this bound holds for all $x\in[0,1]$, the sum is empty for $x\in(1-\frac{1}{q},1]$ and trivially equal to zero. For $x\geq\frac{1}{2}$, an upper bound on the integral is given by
\begin{align}
 & \int_{\left\lceil qx\right\rceil /q}^{1}e^{-q\beta D(y||x)}\dd y\leq\int_{x}^{1}e^{-q\beta D(y||x)}\dd y\nonumber \\
 & \stackrel{(a)}{\leq}\int_{x}^{1}\exp\left(\frac{-q\beta(y-x)^{2}}{2(x(1-x)+(1-2x)(y-x)/3)}\right)\dd y\nonumber \\
 & \quad\stackrel{(b)}{\leq}\int_{x}^{\infty}\exp\left(-\frac{q\beta(y-x)^{2}}{2x(1-x)}\right)\dd y\nonumber \\
 & \quad\quad=\sqrt{\frac{\pi x(1-x)}{2q\beta}},\label{eq:upper_sum_int_ub}
\end{align}
where $(a)$ follows from Lemma~\ref{lem:Simple_KL_bin_bound} and $(b)$ holds because $(1-2x)(y-x)\leq0$ for $y\geq x\geq\frac{1}{2}$.

Now, we consider the lower sum in~\eqref{eq:TV_sum_decomp}. Similarly, for $i\in\{0,\ldots,\left\lceil qx\right\rceil -2\}$, \eqref{eq:Binomial_KL2}~shows that
\[
\psi_{i}(x)=1-\Pr\left(\mathrm{Bin}(q,x)\leq i\right)\geq1-e^{-qD(\frac{i}{q}||x)}.
\]
For $i\in\{0,1,\left\lceil qx\right\rceil -2\}$, we have $V(1-\psi_{i}(x))\leq(1-\psi_{i}(x))^{\beta}\leq e^{-q\beta D(\frac{i}{q}||x)}$ and thus
\begin{align}
\frac{1}{q}\sum_{i=0}^{\left\lceil qx\right\rceil -2} & V\left(\psi_{i}(x)\right)=\frac{1}{q}\sum_{i=0}^{\left\lceil qx\right\rceil -2}V\left(1-\psi_{i}(x)\right)\nonumber \\
 & \leq\frac{1}{q}\sum_{i=0}^{\left\lceil qx\right\rceil -2}e^{-q\beta D(\frac{i}{q}||x)}\nonumber \\
 & \stackrel{(a)}{\leq}\frac{1}{q}\sum_{i=0}^{\left\lceil qx\right\rceil -2}\int_{0}^{1}e^{-q\beta D(\frac{i+z}{q}||x)}\dd z\nonumber \\
 & \stackrel{(b)}{=}\int_{0}^{(\left\lceil qx\right\rceil -1)/q}e^{-q\beta D(y||x)}\dd y,\label{eq:Tq_bound_lower}
\end{align}
where $e^{-q\beta D(\frac{i}{q}||x)}\leq\int_{0}^{1}e^{-q\beta D(\frac{i+z}{q}||x)}\dd z$ holds in $(a)$ because $e^{-q\beta D(\frac{i+z}{q}||x)}$ is increasing in $z$ for $z\in[0,1]$ and $i+1\leq qx$. Also, $(b)$ follows from grouping terms into one integral and changing the variable of integration.

The expression in~\eqref{eq:Tq_bound_lower} can be upper bounded using the decomposition
\begin{align}
\int_{0}^{(\left\lceil qx\right\rceil -1)/q} & e^{-q\beta D(y||x)}\dd y\leq\int_{0}^{x}e^{-q\beta D(y||x)}\dd y\nonumber \\
 & \leq\int_{0}^{2x-1}e^{-q\beta D(y||x)}\dd y+\int_{2x-1}^{x}e^{-q\beta D(y||x)}\dd y.\label{eq:Tq_bound_lower_split}
\end{align}
The first term in~\eqref{eq:Tq_bound_lower_split} can be upper bounded with 
\begin{align}
 & \int_{2x-1}^{x}e^{-q\beta D(y||x)}dy\nonumber \\
 & \stackrel{(a)}{\leq}\!\!\int_{2x-1}^{x}\!\!\!\!\!\exp\left(\frac{-q\beta(y-x)^{2}}{2(x(1-x)+(2x-1)(x-y)/3)}\right)\dd y\nonumber \\
 & \stackrel{(b)}{\leq}\!\!\int_{2x-1}^{x}\!\!\!\exp\left(\frac{-q\beta(y-x)^{2}}{2((1-x)+(1-x)/3)}\right)\dd y\nonumber \\
 & =\sqrt{\frac{2\pi(1-x)}{3q\beta}}\text{erf}\left(\sqrt{\frac{3\beta q(1-x)}{8}}\right)\nonumber \\
 & \stackrel{(c)}{\leq}\sqrt{\frac{4\pi x(1-x)}{3q\beta}},\label{eq:lowersum1_final_ub}
\end{align}
where $(a)$ follows from Lemma~\ref{lem:Simple_KL_bin_bound}, $(b)$ holds because $x(1-x)\leq1-x$ and $(2x-1)(x-y)\leq1-x$ for $y\geq2x-1$ and $x\geq\frac{1}{2}$, and $(c)$ follows from $2x\geq1$ for $x\geq\frac{1}{2}$. The second term in~\eqref{eq:Tq_bound_lower_split} can be upper bounded with 
\begin{align}
 & \int_{0}^{2x-1}e^{-q\beta D(y||x)}\dd y\nonumber \\
 & \stackrel{(a)}{\leq}\!\!\int_{0}^{2x-1}\!\!\!\!\!\!\!\!\exp\left(-q\beta\left((y-x)+(1-y)\ln\frac{1-y}{1-x}\right)\right)\dd y\nonumber \\
\,\nonumber \\
 & \stackrel{(b)}{=}\!\!\int_{\frac{1}{1-x}}^{2}\!\!\!\exp\bigg(-q\beta\bigg((1-z(1-x)-x)\nonumber \\
 & \quad\quad\quad\quad\quad\quad+z(1-x)\ln\frac{z(1-x)}{1-x}\bigg)\bigg)(x\!-\!1)\dd z\nonumber \\
 & =(1\!-\!x)\int_{2}^{\frac{1}{1-x}}\!\!\exp\left(q\beta(x-1)\left((1-z)+z\ln z\right)\right)\dd z\nonumber \\
 & \stackrel{(c)}{\leq}(1\!-\!x)\int_{2}^{\frac{1}{1-x}}\!\!\exp\left(\frac{1}{2}q\beta(x-1)(z-1)^{2}\right)\dd z\nonumber \\
 & =(1\!-\!x)\sqrt{\frac{\pi}{2q\beta(1-x)}}\text{erf}\sqrt{\frac{q\beta(1-x)(z-1)^{2}}{2}}\bigg|_{z=2}^{\frac{1}{1-x}}\nonumber \\
 & \stackrel{(d)}{\leq}(1-x)\sqrt{\frac{\pi}{2q\beta(1-x)}}\nonumber \\
 & \stackrel{(e)}{\leq}\sqrt{\frac{\pi x(1-x)}{q\beta}},\label{eq:lowersum2_final_ub}
\end{align}
where $(a)$ follows from Lemma~\ref{lem:KL_Poisson}, $(b)$ is given by the change of variables $y\mapsto1-z(1-x)$, $(c)$ holds because $1-z+z\ln z\geq\frac{1}{2}(z-1)^{2}$ for $z\in[0,1]$, $(d)$ follows from $\mbox{erf}(b)-\mbox{erf}(a)\leq1$ for $b\geq a\geq0$, and $(e)$ holds because $2x\geq1$ for $x\geq\frac{1}{2}$.

Now, we combine Lemma~\ref{lem:g_psi_middle_term} with~\eqref{eq:upper_sum_int_ub}, \eqref{eq:lowersum1_final_ub}, and \eqref{eq:lowersum2_final_ub} to see that 
\begin{align}
 & (T_{q}V)(x)=\frac{1}{q}\Bigg(\sum_{i=0}^{\left\lceil qx\right\rceil -2}V\left(\psi_{i}(x)\right)+V\left(\psi_{\left\lceil qx\right\rceil -1}(x)\right)\nonumber \\
 & \quad\quad\quad\quad\quad\quad\quad+\sum_{i=\left\lceil qx\right\rceil }^{q-1}V\left(\psi_{i}(x)\right)\Bigg)\nonumber \\
 & \leq\int_{0}^{(\left\lceil qx\right\rceil -1)/q}e^{-q\beta D(y||x)}\dd y+\frac{1}{q}V\left(\psi_{\left\lceil qx\right\rceil -1}(x)\right)\nonumber \\
 & \quad\quad\quad+\int_{\left\lceil qx\right\rceil /q}^{1}e^{-q\beta D(y||x)}\dd y\nonumber \\
 & \leq\sqrt{\frac{\pi x(1-x)}{q\beta}}+\sqrt{\frac{4\pi x(1-x)}{3q\beta}}\nonumber \\
 & \quad\quad\quad+\frac{\left(2x(1-x)\right)^{\beta}}{\sqrt{2q}}+\sqrt{\frac{\pi x(1-x)}{2q\beta}}\nonumber \\
 & \leq\frac{\left(2x(1-x)\right)^{\beta}}{\sqrt{2q}}+A\sqrt{\frac{x(1-x)}{q\beta}},\label{eq:TqV_ub1}
\end{align}
where $A=\sqrt{\pi}+\sqrt{\frac{4\pi}{3}}+\sqrt{\frac{\pi}{2}}$. Combining Definition~\ref{def:lambda_q_b} with~\eqref{eq:TqV_ub1}, we see that 
\begin{align*}
\lambda_{q,\beta} & \triangleq\sup_{x\in(0,1)}\frac{(T_{q}V)(x)}{V(x)}\\
\leq & \sup_{x\in(0,1)}\left(\frac{2^{\beta}}{\sqrt{2q}}+\frac{A}{\sqrt{q\beta}}\left(x(1-x)\right)^{1/2-\beta}\right)\\
\stackrel{(a)}{\leq} & \frac{2^{\beta}}{\sqrt{2q}}+A\frac{1}{\sqrt{q\beta}}\left(\frac{1}{4}\right)^{\frac{1}{2}-\beta}\\
\leq & \frac{1}{\sqrt{q\beta}}\left(\frac{1}{4}\right)^{\frac{1}{2}-\beta}\left(A+\frac{2^{\beta}\sqrt{\beta}}{\sqrt{2}\left(\frac{1}{4}\right)^{\frac{1}{2}-\beta}}\right)\\
 & \stackrel{(b)}{\leq}\frac{6}{\sqrt{q\beta}}\left(\frac{1}{4}\right)^{\frac{1}{2}-\beta},
\end{align*}
where $(a)$ holds because the supremum is achieved at $x=\frac{1}{2}$ and $(b)$ holds because
\[
A+\frac{2^{\beta}\sqrt{\beta}}{\sqrt{2}\left(\frac{1}{4}\right)^{\frac{1}{2}-\beta}}\leq6
\]
for $\beta\in(0,\frac{1}{2}]$.

\section{\label{sec:fixed_alphabet} Fixed-Alphabet Erasure Channels with\protect \\
Large Polarizing Matrices}

The results in the previous section show that one can approach the optimal scaling rate for $q$-ary erasure channels with $q\times q$ polarizing matrices. It does not, however, allow one to separate the effect of alphabet size and transform size. In this section, we explicitly consider this distinction by constructing a polar code based on $m\times m$ polarizing matrices over $\mathbb{F}_{q}$ and focus on the scaling as $m$ increases. The polar decoder at each stage is based on APP decoding the $m$ nested subcodes associated with successive cancellation decoding. The rate of polarization in $m$ is investigated numerically and a formula for the general case is conjectured.

Let $GL(m,\mathbb{F}_{q})$ denote the general linear group of invertible $m\times m$ matrices over $\mathbb{F}_{q}$. For a fixed $G'\in GL(m,\mathbb{F}_{q})$, let $\varphi_{i}(x;G')$ be the erasure rate (under APP decoding) of the first input symbol of the $(n,n-i)$ linear encoder defined by the bottom $n-i$ rows of $G'$, assuming the output symbols are erased with probability $x$. Consider the length $N=m^{n}$ polar code defined by the polar transform $G_{0}^{\otimes n}$ for a fixed $G_{0}\in GL(m,\mathbb{F}_{q})$. During each stage of decoding, effective channels with erasure rate $x$ are transformed into $m$ different effective channels with erasure rates $\varphi_{i}(x;G_{0})$ for $i\in\mathcal{M}\triangleq\left\{ 0,1,\ldots,m-1\right\} $. Let the random variable $X_{n}$ denote the erasure probability of a randomly chosen effective channel after $n$ levels of polarization~\cite{Hassani-it13,Mondelli-arxiv15}. Then, the sequence $X_{n}$ is a Markov chain on the compact state space $\mathcal{X}=[0,1]$ with transition kernel 
\begin{align*}
\Pr\Big( & X_{n}\,=x_{n}\,\Big|\,(X_{0},\ldots,X_{n-1})=(x_{0},\ldots,x_{n-1})\Big)\\
 & =\Pr\left(X_{n}=x_{n}\midb |X_{n-1}=x_{n-1}\right)\\
 & \quad=\frac{1}{m}\left|\left\{ i\in\mathcal{M}\,|\,x_{n}=\varphi_{i}(x_{n-1};G_{0})\right\} \right|.
\end{align*}

Similar to previous examples, one can analyze this Markov chain by focusing on the cost function $g_{n}(x)\triangleq\expt[g_{0}(X_{n})|X_{0}=x]$ generated by $g_{0}\in C(\mathcal{X})$. The one-step update for $g_{n}$ is given by the linear operator $T_{m}:C(\mathcal{X})\to C(\mathcal{X})$ where
\[
g_{n+1}(x)=(T_{m}g_{n})(x)\triangleq\frac{1}{m}\sum_{i=0}^{m-1}g_{n}\left(\varphi_{i}(x;G_{0})\right).
\]
Numerical results show that the length-16 BCH kernel from~\cite{Korada-it10*2} has a better rate of polarization than \arikan's binary polar codes. If one chooses $g_{0}(x)=(x(1-x))^{0.6}$ for this kernel, then one finds that 
\[
\lambda_{BCH16}=\sup_{x\in[0,1]}\frac{g_{1}(x)}{g_{0}(x)}\approx0.4508.
\]
This is better than the mixing value $\lambda_{2}^{4}\approx(0.827)^{4}\approx0.4677$ given by 4 stages of binary polarization. Indeed, this is quite related to the error exponent of polar codes. Computing this quantity for any fixed $m\times m$ kernel requires summing over all $2^{m}$ subsets. Thus, it is computationally challenging to extend these results much beyond length 16 (e.g., to length 32).

For the sake of analysis, we consider an inhomogeneous polar code where the polarizing matrix (or kernel) at each stage is different. Thus, the overall length-$N=m^{n}$ polar transform is defined by the generator matrix
\[
G=G_{1}\otimes G_{2}\otimes\cdots\otimes G_{n},
\]
where the sequence $G_{1},G_{2},\ldots,G_{n}$ is drawn i.i.d. from the general linear group $GL(m,\mathbb{F}_{q})$ of invertible $m\times m$ matrices over $\mathbb{F}_{q}$. We note that unconstrained i.i.d. random matrices cannot be used in this construction because the preservation of mutual information required by polarization fails if the kernel is not invertible. Let the random variable $X_{n}$ denote the channel erasure probability for a randomly chosen effective channel after $n$ levels of polarization~\cite{Hassani-it13,Mondelli-arxiv15}. Then, the sequence $X_{n}$ is a Markov chain on the compact state space $\mathcal{X}=[0,1]$ with transition kernel
\begin{align*}
\Pr & \left(X_{n}=x_{n}\midb |X_{n-1}=x_{n-1}\right)=\frac{1}{m\left|GL(m,\mathbb{F}_{q})\right|}\\
 & \cdot\left|\left\{ (i,G_{n})\in\mathcal{M}\times GL(m,\mathbb{F}_{q})\,|\,x_{n}=\varphi_{i}(x_{n-1};G_{n})\right\} \right|.
\end{align*}
Since this probability also includes the randomness in the kernel selection, it is worth noting that the conditional fraction of unpolarized channels, $\Pr\left(X_{n}\in[\eta,1-\eta]\,\big|\,G_{1},\ldots,G_{n}\right)$, is a random variable. Thus, the Markov inequality implies that 
\begin{align*}
\Pr\Big(\Big\{\Pr\big(X_{n}\in[\eta,1-\eta] & \,\big|\,G_{1},\ldots,G_{n}\big)>\delta\Big\}\Big)\\
 & <\frac{1}{\delta}\Pr\left(X_{n}\in[\eta,1-\eta]\right).
\end{align*}
Hence, choosing $\delta=\Pr\big(X_{n}\in[\eta,1-\eta]\big)$ shows that there is at least one code with, at most, a fraction $\Pr\big(X_{n}\in[\eta,1-\eta]\big)$ of unpolarized channels.

Similar to previous examples, one can analyze this Markov chain by focusing on the cost function $g_{n}(x)\triangleq\expt[g_{0}(X_{n})|X_{0}=x]$ generated by $g_{0}\in C(\mathcal{X})$. The one-step update for $g_{n}$ is given by the linear operator $T_{m}:C(\mathcal{X})\to C(\mathcal{X})$ where
\begin{align*}
g_{n+1}(x) & =(T_{m}g_{n})(x)\triangleq\frac{1}{\left|GL(m,\mathbb{F}_{q})\right|}\\
 & \quad\cdot\sum_{G_{n}\in GL(m,\mathbb{F}_{q})}\frac{1}{m}\sum_{i=0}^{m-1}g_{n}\left(\varphi_{i}(x;G_{n})\right).
\end{align*}
If one chooses $g_{0}(x)$ to be concave, then one also gets the upper bound
\begin{align*}
g_{1}(x) & =\frac{1}{\left|GL(m,\mathbb{F}_{q})\right|}\sum_{G_{1}\in GL(m,\mathbb{F}_{q})}\frac{1}{m}\sum_{i=0}^{m-1}g_{0}\left(\varphi_{i}(x;G_{1})\right)\\
 & \leq\frac{1}{m}\sum_{i=0}^{m-1}g_{0}\left(\sum_{G_{1}\in GL(m,\mathbb{F}_{q})}\frac{\varphi_{i}(x;G_{1})}{\left|GL(m,\mathbb{F}_{q})\right|}\right)\\
 & \triangleq\overline{g}_{1}(x).
\end{align*}
Continuing recursively, one can define
\begin{align*}
\overline{g}_{n+1}(x) & \triangleq\frac{1}{m}\sum_{i=0}^{m-1}\overline{g}_{n}\left(\sum_{G_{n+1}\in GL(m,\mathbb{F}_{q})}\frac{\varphi_{i}(x;G_{n+1})}{\left|GL(m,\mathbb{F}_{q})\right|}\right)\\
 & =\frac{1}{m}\sum_{i=0}^{m-1}\overline{g}_{n}\left(\overline{\varphi}_{i}(x)\right),
\end{align*}
where $\overline{\varphi}_{i}(x)$ is the average of $\varphi_{i}(x;G')$ over all $G'\in GL(m,\mathbb{F}_{q})$. Similarly, if $g_{n}(x)\leq\overline{g}_{n}(x)$ and $\overline{g}_{n}(x)$ is concave, then one can write
\begin{align*}
g_{n+1}(x) & \leq\frac{1}{m}\sum_{i=0}^{m-1}g_{n}\left(\overline{\varphi}_{i}(x)\right)\\
 & \leq\frac{1}{m}\sum_{i=0}^{m-1}\overline{g}_{n}\left(\overline{\varphi}_{i}(x)\right)=\overline{g}_{n+1}(x)
\end{align*}
to derive an inductive upper bound. To complete the analysis, we need a formula for $\overline{\varphi}_{i}(x)$ and a concave cost function $g_{0}(x)$ such that $\overline{g}_{n}(x)$ is a sequence of concave functions. A closed-form expression for $\overline{\varphi}_{i}(x)$, which can be evaluated using $O(m^{4})$ real operations, is derived in Appendix~\ref{sec:full_rank}. 

For $q=2$ and $m=16,32,64$, one can choose $g_{0}(x)=(x(1-x))^{0.35}$ and observe that $\overline{g}_{1}(x)$ is concave. Moreover, the constants 
\[
\lambda_{m}=\sup_{x\in[0,1]}\frac{\overline{g}_{1}(x)}{g_{0}(x)}
\]
are given by $(\lambda_{16},\lambda_{32},\lambda_{64})\approx(0.6729,0.4558,0.2880)$. Comparing with binary polar codes, where the mixing rate is roughly $0.827^{\log_{2}m}$, one finds that $\lambda_{64}\approx0.2880\leq0.3199\approx0.827{}^{6}$. Thus, this predicts that polar codes based on random invertible $64\times64$ binary polarizing kernels will achieve a better scaling rate that \arikan's binary polar codes. This statement is not rigorous, however, because we have not shown that $\overline{g}_{n}(x)$ remains concave for all $n$. This leads to the conjecture.
\begin{conjecture}
For any prime-power $q\geq2$, any integer $m\geq2$, and any $\beta\in(0,1)$, let $g_{0}(x)=(x(1-x))^{\beta}$. Then, $\overline{g}_{n}(x)$ is concave on $[0,1]$ for all $n$.
\end{conjecture}
If this conjecture is true, then this sequence of inhomogeneous polar codes can be rigorously characterized in terms of $\lambda_{m}$. Additionally, our next conjecture is sufficient to imply that there is a sequence of inhomogeneous polar codes that achieves near-optimal scaling with fixed $q$ as $m$ increases.
\begin{conjecture}
For any prime-power $q\geq2$ and any $\beta\in(0,\frac{1}{2}]$, let $g_{0}(x)=(x(1-x))^{\beta}$. Then, 
\[
\lim_{m\to\infty}\frac{1}{\ln m}\ln\lambda_{m}=\lim_{m\to\infty}\frac{1}{\ln m}\ln\sup_{x\in[0,1]}\frac{\overline{g}_{1}(x)}{g_{0}(x)}=-\frac{1}{2}.
\]
\end{conjecture}
If these conjectures are both true, then Corollary~\ref{cor:lyapunov} implies that, for any $\gamma>0$ and $\beta\in(0,\frac{1}{2}]$, we have 
\begin{align*}
\Pr\left(X_{n}\geq N^{-\gamma}\,|\,X_{0}=x\right) & \leq N^{\gamma\beta}e^{n\ln m\frac{\ln\lambda_{m}}{\ln m}}+\frac{x}{1-N^{-\gamma}}\\
 & =N^{\gamma\beta-\frac{1}{2}+o_{m}(1)}+\frac{x}{1-N^{-\gamma}}.
\end{align*}

\section{Conclusion}

In this paper, we first investigate the relationship between the blocklength and the gap to capacity for the $q$-ary Reed-Solomon polar codes introduced by Mori and Tanaka. These codes have length $N=q^{n}$, where $n$ is the number of steps in the polarization process. When one of these codes is transmitted over a $q$-ary erasure channel with erasure probability $\epsilon$, its effective channels are $q$-ary erasure channels and their erasure rate satisfy a closed-form recursion. By analyzing this recursion, we show that , for any $\gamma>0$ and $\delta>0$, there is a $q_{0}$ such that, for all $q\geq q_{0}$, the fraction of effective channels with erasure rate at most $O(N^{-\gamma})$ is at least $1-\epsilon-O(N^{-1/2+\delta})$. Thus, the gap to capacity scales at a rate very close to the optimal rate of $O(N^{-1/2})$.

In the second part of this paper, a similar analysis is also considered for $q$-ary polar codes with $m$ by $m$ polarizing matrices. To prove near-optimal scaling for polar codes with fixed $q$ as $m$ increases, however, two technical obstacles remain. Thus, we conclude by stating two concrete mathematical conjectures that, if proven, would imply near-optimal scaling.

These results naturally suggest two interesting open questions. First, can one prove that $q$-ary inhomogeneous polar codes with random $m\times m$ polarization kernels achieve near-optimal scaling on the $q$-ary erasure channel with fixed $q$ as $m$ increases? Second, can this result be extended to noisy $q$-ary (or even binary) channels? 

\appendices

\section{\label{sec:full_rank} The Polar Erasure Rate for\protect \\
Random Full-Rank Matrices}

Let $G$ be an $m\times m$ full-rank matrix and let $G^{(i)}$ be the $(m-i)\times m$ submatrix formed by removing the first $i$ rows from $G$. Then, $G^{(i)}$ has rank $m-i$ (i.e., full rank) because removing a row from a matrix reduces its rank by at most 1. Recall that, during polar decoding, one decodes the sequence of full-rank generator matrix codes defined by $G^{(i)}$. Let $\mathcal{S}$ be the set of indices of correctly received output bits and let $G_{\mathcal{S}}^{(i)}$ be the $(m-i)\times|\mathcal{S}|$ submatrix of $G^{(i)}$ containing only columns whose indices are in $\mathcal{S}$. For any fixed $\mathcal{S}$, the $i$-th input bit can be recovered during the $i$-th decoding step iff the $i$-th input bit can be written as a linear combination of the correctly received bits. In this setup, this occurs iff there is a vector $v$ such that $G_{\mathcal{S}}^{(i)}v=e_{1},$ where $e_{j}$ is the unit column vector with a one in the $j$-th position. In terms of matrix ranks, this is equivalent to the condition
\[
\rank\big(G_{\mathcal{S}}^{(i)}\big)=\rank\big([e_{1}\;G_{\mathcal{S}}^{(i)}]\big).
\]
Since $\rank\big([e_{1}\;G_{\mathcal{S}}^{(i)}]\big)\leq\rank\big(G_{\mathcal{S}}^{(i)}\big)+1$, it follows that
\[
E^{(i)}(\mathcal{S})\triangleq\rank\big([e_{1}\;G_{\mathcal{S}}^{(i)}]\big)-\rank\big(G_{\mathcal{S}}^{(i)}\big)
\]
is the erasure indicator function for the $i$-th step of decoding when $\mathcal{S}$ is the set of correctly received bits.

For a random $G$ matrix, we observe that
\[
\expt\left[E^{(i)}(\mathcal{S})\right]=\Pr\left(e_{1}\notin\mbox{colspace}\big(G_{\mathcal{S}}^{(i)}\big)\right)
\]
because the rank is increased by including the column $e_{1}$ iff $e_{1}$ is not in the column space. If $G\in\mathbb{F}_{q}^{m\times m}$ is a uniform random full-rank matrix, then $G^{(i)}\in\mathbb{F}_{q}^{(m-i)\times m}$ is also uniform random full-rank matrix. Since the $G^{(i)}$ ensemble is invariant under column permutations, it follows $\Pr\left(e_{1}\notin\mbox{colspace}\big(G_{\mathcal{S}}^{(i)}\big)\right)$ only depends on the cardinality of $\mathcal{S}$ and not on the individual elements. Hence, we define
\[
\rho(m,i,|\mathcal{S}|,q)\triangleq\Pr\left(e_{1}\notin\mbox{colspace}\big(G_{\mathcal{S}}^{(i)}\big)\right)
\]
for this ensemble.

Now, we assume that each output bit is erased independently with probability $x$ and let $\overline{\varphi}_{i}(x)$ be the erasure probability of the $i$-th input bit during the $i$-th step of polar decoding averaged over the ensemble of full-rank polarizing matrices. In this case, we find that
\begin{align*}
\overline{\varphi}_{i}(x) & \triangleq\expt\left[\sum_{\mathcal{S}\subseteq[m]}x^{m-|\mathcal{S}|}(1-x)^{|\mathcal{S}|}E^{(i)}(\mathcal{S})\right]\\
 & =\sum_{\mathcal{S}\subseteq[m]}x^{m-|\mathcal{S}|}(1-x)^{|\mathcal{S}|}\Pr\left(e_{1}\notin\mbox{colspace}\left(G_{\mathcal{S}}^{(i)}\right)\right)\\
 & =\sum_{d=0}^{m}\binom{m}{d}x^{m-d}(1-x)^{d}\rho(m,i,d,q).
\end{align*}
A closed-form expression for $\rho(m,i,|\mathcal{S}|,q)$ is derived below. 

It is well-known~(e.g., see \cite{Fisher-amm66}) that, for a uniform random matrix $G'\in\mathbb{F}_{q}^{k\times d}$, the rank satisfies
\[
\Pr\left(\rank(G')=j\right)=\phi(j,d,q){k \brack j}_{q}q^{-kd},
\]
where the number of sets of $j$ linearly independent vectors of length $i$ is given by
\[
\phi(j,i,q)\triangleq\prod_{l=0}^{j-1}(q^{i}-q^{l})
\]
and the number of $j$-dimensional subspaces of $\mathbb{F}_{q}^{k}$ equals the Gaussian binomial coefficient
\[
{k \brack j}_{q}\triangleq\prod_{l=0}^{j-1}\frac{q^{k}-q^{l}}{q^{j}-q^{l}}.
\]
For fixed $\mathcal{S}$, consider the submatrix $G_{\mathcal{S}}'$ of a uniform random matrix $G'\in\mathbb{F}_{q}^{k\times m}$ and define $\mathcal{S}^{c}\triangleq\left\{ 1,2,\ldots,m\right\} \backslash\mathcal{S}$. One can compute the joint rank distribution of $G_{\mathcal{S}}'$ and $G'$ using the fact that 

\begin{align*}
\theta & (m,k,r,j,q)\triangleq\Pr\left(\rank\big(G_{\mathcal{S}}'\big)=j,\,\rank\big(G'\big)=r\right)\\
 & =\Pr\left(\rank\big(G_{\mathcal{S}}'\big)=j\right)\Pr\left(\rank\big([G_{\mathcal{S}}'\;G_{\mathcal{S}^{c}}']\big)=r\big|\rank\big(G_{\mathcal{S}}'\big)=j\right)\\
 & =\Pr\left(\rank\big(G_{\mathcal{S}}'\big)=j\right)\sum_{l=0}^{r}\Pr\left(\rank\big(G_{\mathcal{S}^{c}}'\big)=l\right)\\
 & \quad\quad\cdot\Pr\left(\rank\big([G_{\mathcal{S}}'\;G_{\mathcal{S}^{c}}']\big)=r\big|\rank\big(G_{\mathcal{S}}'\big)=j,\rank\big(G_{\mathcal{S}^{c}}'\big)=l\right)\\
 & =\Pr\left(\rank\big(G_{\mathcal{S}}'\big)=j\right)\sum_{l=0}^{r}\Pr\left(\rank\big(G_{\mathcal{S}^{c}}'\big)=l\right)\\
 & \quad\quad\cdot\Pr\left(\dim\left(W\oplus W'\right)=r\,|\,\dim(W)=j,\dim(W')=l\right),
\end{align*}
where $W$ and $W'$ are the independent uniform random column-spaces of $G_{\mathcal{S}}'$ and $G_{\mathcal{S}^{c}}'$. For independent uniform random subspaces $W,W'$ of $\mathbb{F}_{q}^{k}$, it is shown in~\cite{Rathi-iee05} that 
\begin{align*}
\Pr & \left(\dim\left(W\oplus W'\right)=r\,|\,\dim(W)=j,\dim(W')=l\right)\\
 & \quad\quad=q^{(r-j)(r-l)}{j \brack r-l}_{q}{k-j \brack k-r}_{q}\bigg/{k \brack l}_{q}.
\end{align*}
Therefore, we can write
\begin{align}
\Pr & \left(\rank\big(G_{\mathcal{S}}'\big)=j,\,\rank\big(G'\big)=r\right)=\phi(j,d,q){k \brack j}_{q}q^{-kd}\nonumber \\
 & \;\cdot\sum_{l=0}^{r}\phi(l,m-d,q)q^{-k(m-d)}q^{(r-j)(r-l)}{j \brack r-l}_{q}{k-j \brack k-r}_{q}.\label{eq:joint_rank}
\end{align}

Now, we have the tools to complete the main derivation. Let $\mathcal{S}\subseteq[m]$ be an arbitrary subset satisfying $|\mathcal{S}|=d$ and write
\begin{align*}
\rho & (m,i,d,q)=\Pr\left(e_{1}\notin\mbox{colspace}\big(G_{\mathcal{S}}^{(i)}\big)\right)\\
 & =\sum_{j=0}^{m-i}\Pr\left(\rank\big(G_{\mathcal{S}}^{(i)}\big)=j\right)\\
 & \quad\quad\cdot\Pr\left(e_{1}\notin\mbox{colspace}\big(G_{\mathcal{S}}^{(i)}\big)\,\big|\,\rank\big(G_{\mathcal{S}}^{(i)}\big)=j\right)\\
 & \stackrel{(a)}{=}\sum_{j=0}^{m-i}\Pr\left(\rank\big(G_{\mathcal{S}}'\big)=j\,\big|\,\rank\big(G'\big)=m-i\right)\\
 & \quad\quad\cdot\Pr\left(e_{1}\notin\mbox{colspace}\big(G_{\mathcal{S}}'\big)\,\big|\,\rank\big(G_{\mathcal{S}}'\big)=j\right)\\
 & \stackrel{(b)}{=}\sum_{j=0}^{m-i}\Pr\left(\rank\big(G_{\mathcal{S}}'\big)=j\,\big|\,\rank\big(G'\big)=m-i\right)\frac{q^{m-i}-q^{j}}{q^{m-i}-1}\\
 & =\sum_{j=0}^{m-i}\left(\frac{q^{m-i}-q^{j}}{q^{m-i}-1}\right)\frac{\Pr\left(\rank\big(G_{\mathcal{S}}'\big)=j,\,\rank\big(G'\big)=m-i\right)}{\Pr\left(\rank\big(G'\big)=m-i\right)}\\
 & \stackrel{(c)}{=}\sum_{j=0}^{m-i}\left(\frac{q^{m-i}-q^{j}}{q^{m-i}-1}\right)\frac{\phi(j,d,q){m-i \brack j}_{q}q^{-(m-i)d}}{\phi(m-i,m,q)q^{-(m-i)m}}q^{-(m-i)(m-d)}\\
 & \quad\quad\cdot\sum_{l=0}^{m-i}\phi(l,m-d,q)q^{(m-i-j)(m-i-l)}{j \brack m-i-l}_{q}\\
 & =\sum_{j=0}^{m-i}\left(\frac{q^{m-i}-q^{j}}{q^{m-i}-1}\right)\frac{\phi(j,d,q){m-i \brack j}_{q}}{\phi(m-i,m,q)}\\
 & \quad\quad\cdot\sum_{l=0}^{m-i}\phi(l,m-d,q)q^{(m-i-j)(m-i-l)}{j \brack m-i-l}_{q},
\end{align*}
where $(a)$ follows from the fact that a uniform random matrix $G'\in\mathbb{F}_{q}^{(m-i)\times m}$ has the same distribution as $G^{(i)}$ conditioned on the event that $\rank\big(G_{\mathcal{S}}'\big)=m-i$, $(b)$ follows from the fact that the $G^{(i)}$ ensemble is invariant under left-multiplication by a random invertible matrix and $(c)$ follows from~\eqref{eq:joint_rank}. 

Let $G^{(i)}$ be a random full-rank $(m-i)\times m$ matrix. Then, the choice of $G^{(i)}$ can be separated into the choice of a uniform random subspace, $\mathcal{C}\subseteq\mathbb{F}_{q}^{m}$, and a uniform random basis for that subspace. Let $H\in\mathbb{F}_{q}^{i\times m}$ be a uniform random basis for the dual space $\mathcal{C}^{\perp}$. Then, $H$ is a uniform random parity-check matrix for the code generated by $G^{(i)}$ (i.e., $G^{(i)}H^{T}=0_{(m-i)\times i})$, where $0_{a\times b}$ denotes an $a\times b$ all-zero matrix. Since the choice of $H$ can also be separated into the choice of a uniform random subspace, $\mathcal{C}^{\perp}$, and a uniform random basis, it follows that the marginal distribution of $H$ (i.e., averaged over the choice of $G^{(i)}$) is uniform over the set of full-rank $i\times m$ matrices. Similarly, the conditional distribution of $G^{(i)}$ given $H$ is uniform over the set of bases for $\big(\mathcal{C}^{\perp}\big)^{\perp}=\mathcal{C}$. 

For any full-rank generator matrix $G^{(i)}\in\mathbb{F}_{q}^{(m-i)\times m}$, one can define the augmented generator matrix $\tilde{G}=[G^{(i)}\;e_{1}]$. For the $(m+1,m-i)$ linear code generated by $\tilde{G}$, the recovery of the first information bit (of either code) is equivalent to the extrinsic recovery of the last code bit of the augmented code. Let $\tilde{H}$ be a $(i+1)\times(m+1)$ parity-check matrix for the augmented code (i.e., $\tilde{G}\tilde{H}^{T}=0_{(m-i)\times(i+1)})$. We construct a (non-uniform) random $\tilde{H}$ matrix on the same probability space by defining 
\[
\tilde{H}\triangleq\begin{bmatrix}v^{T} & 1\\
H & 0_{m}
\end{bmatrix},
\]
where $v$ is a uniform random solution to $G^{(i)}v=-e_{1}$. Based on this definition, one can verify that
\[
\tilde{G}\tilde{H}^{T}=[G^{(i)}\;e_{1}]\begin{bmatrix}v & H^{T}\\
1 & 0_{m}^{T}
\end{bmatrix}=\begin{bmatrix}0_{m-i} & 0_{(m-i)\times(i+1)}\end{bmatrix}.
\]

Now, we will show that the matrix 
\[
\tilde{H}'\triangleq\begin{bmatrix}v\\
H
\end{bmatrix}
\]
is a uniform random $(i+1)\times m$ full-rank matrix. To do this, we first recall that $H$ is a uniform random $i\times m$ full-rank matrix and, conditioned on $H$, the rows of $G^{(i)}$ form a uniform random basis for the null space of $H$. This means that we can write $G^{(i)}=BG'$, where $B$ is uniform random element of $GL(m-i,\mathbb{F}_{q})$ and $G'$ is any basis for the null space of $H$. Now, the vector $v$ is chosen to be a uniform random solution of the system $G^{(i)}v=-e_{1}$ which can be rewritten as $Gv=-B^{-1}e_{1}$. Since $-B^{-1}e_{1}$ is a uniform random non-zero vector, the vector $v$ is distributed uniformly over the set $A$ of vectors such that $Gv\neq0$. Then, $A=\mathbb{F}_{q}^{m}\backslash\text{rowspace}(H)$ because the null space of $G$ equals the row space of $H$. Moreover, $A$ is exactly equal to the set of row vectors that can be used to extend $H$ to a full-rank $(i+1)\times m$ matrix. Thus, $\tilde{H}'$ is a uniform random $(i+1)\times m$ full-rank matrix.

Now, we observe that there exists a vector $u$ such that $G_{\mathcal{S}}^{(i)}u_{\mathcal{S}}=e_{1}$ and $u_{\mathcal{S}^{c}}=0$ if and only if $e_{1}\in\mbox{colspace}\big(G_{\mathcal{S}}^{(i)}\big)$. If and only if such a $u$ vector exists, then we have the equivalence
\[
\tilde{H}_{\mathcal{S}^{c}}=\begin{bmatrix}v_{\mathcal{S}^{c}}^{T}\\
H_{\mathcal{S}^{c}}
\end{bmatrix}\sim\begin{bmatrix}u_{\mathcal{S}^{c}}^{T}\\
H_{\mathcal{S}^{c}}
\end{bmatrix}=\begin{bmatrix}0\\
H_{\mathcal{S}^{c}}
\end{bmatrix},
\]
where $\sim$ indicates equivalence under elementary row operations. This step holds because the set of $v$'s satisfying $G^{(i)}v=-e_{1}$ is a coset of the dual code, which is generated by $H$. Since 
\[
\tilde{H}_{\mathcal{S}^{c}}\sim\begin{bmatrix}0\\
H_{\mathcal{S}^{c}}
\end{bmatrix}
\]
 if and only if $e_{1}\notin\mbox{colspace}\big(\tilde{H}_{\mathcal{S}^{c}}\big)$, we find that $e_{1}\in\mbox{colspace}\big(G_{\mathcal{S}}^{(i)}\big)$ if and only if $e_{1}\notin\mbox{colspace}\big(\tilde{H}_{\mathcal{S}^{c}}\big)$.

Based on the above construction, we see that $G_{\mathcal{S}}^{(i)}$ is an $(m-i)\times d$ submatrix of a random full-rank $(m-i)\times m$ matrix and $\tilde{H}_{\mathcal{S}^{c}}$ is an $(i+1)\times(m-d)$ submatrix of a random full-rank $(i+1)\times m$ matrix. Thus,  we find that
\begin{align*}
\rho(m,i,d,q) & =\Pr\left(e_{1}\notin\mbox{colspace}\big(G_{\mathcal{S}}^{(i)}\big)\right)\\
 & =1-\Pr\left(e_{1}\notin\mbox{colspace}\big(\tilde{H}_{\mathcal{S}^{c}}\big)\right)\\
 & =1-\rho(m,m-i-1,m-d,q).
\end{align*}
Using this, we also observe that
\begin{align*}
1 & -\overline{\varphi}_{i}(x)=\sum_{d=0}^{m}\binom{m}{d}(1-x)^{d}x^{m-d}\left(1-\rho(m,i,d,q)\right)\\
 & =\sum_{d=0}^{m}\binom{m}{d}(1-x)^{d}x^{m-d}\rho(m,m-i-1,m-d,q)\\
 & =\sum_{d'=0}^{m}\binom{m}{m-d'}(1-x)^{m-d'}x^{d'}\rho(m,m-i-1,d',q)\\
 & =\overline{\varphi}_{m-i-1}(1-x).
\end{align*}

It can also be observed, numerically, that
\[
\sum_{i=0}^{m-1}\rho(m,i,d,q)=m-d.
\]
Using this, one can check explicitly that the average polar transform preserves the mutual information
\begin{align*}
\frac{1}{m}\sum_{i=1}^{m} & \overline{\varphi}_{i}(x)=\frac{1}{m}\sum_{i=1}^{m}\sum_{d=0}^{m}\binom{m}{d}x^{m-d}(1-x)^{d}\rho(m,i,d,q)\\
 & =\frac{1}{m}\sum_{d=0}^{m}\binom{m}{d}x^{m-d}(1-x)^{d}(m-d)\\
 & =x.
\end{align*}


\end{document}